\documentclass[aps,prd,onecolumn,showkeys,groupedaddress,showpacs,nofootinbib]{revtex4}
\usepackage{epsfig} \usepackage{amsmath} \usepackage{amsfonts}
\usepackage{amssymb} \usepackage{graphicx} \usepackage{colordvi}
\usepackage{psfrag}
 \usepackage{times}
 \usepackage{amsmath, amsthm, amssymb}
 \usepackage{makeidx}

\newtheorem{Proposition}{Proposition}[section]

\newfont{\gotico}{eufm10 scaled\magstephalf}
\newfont{\qvd}{msam10 scaled\magstephalf}

\def\demo{\par\noindent{\sc Proof. }\begingroup}
\def\enddemo{\hskip1em \mbox{\qvd \char3}\endgroup\par\medskip}

\def\de#1/de#2{\frac{\partial {#1}}{\partial {#2}}}
\def\De#1/de#2{\dfrac{\partial {#1}}{\partial {#2}}}

\def\const{{\rm const.}}

\makeindex \makeatletter
\def\widebar{\accentset{{\cc@style\underline{\mskip10mu}}}}
\makeatother

\begin{document}
\def\bib#1{[{\ref{#1}}]}
\title{\bf The Cauchy problem for metric-affine $f(R)$-gravity in presence of perfect-fluid matter}

\author{S. Capozziello$^{1}$ and S. Vignolo$^{2}$ }

\affiliation{$~^{1}$ Dipartimento di Scienze Fisiche,
Universit\`{a} ``Federico II'' di Napoli and INFN Sez. di Napoli,
Compl. Univ. Monte S. Angelo Ed. N, via Cinthia, I- 80126 Napoli
(Italy)}

\affiliation{$^{2}$DIPTEM Sez. Metodi e Modelli Matematici,
Universit\`a di Genova,  Piazzale Kennedy, Pad. D - 16129 Genova
(Italy)}

\date{\today}

\begin{abstract}
The Cauchy problem for metric-affine $f(R)$-gravity {\it \`a la\/} Palatini and with torsion, in presence
of perfect fluid matter acting as  source, is discussed following the well-known Bruhat prescriptions
for General Relativity. The problem results {\it well-formulated}  and {\it well-posed} when the perfect-fluid form of the stress-energy tensor is preserved under conformal transformations. The key role of conservation laws in Jordan and in Einstein frame is also discussed.
\end{abstract}


\maketitle

\section{Introduction}
Several extensions of General Relativity have recently acquired great interest in cosmology and in quantum field theory in order to cure  the shortcomings of  Einstein's theory at ultra-violet and infra-red scales. In particular, such relativistic theories could result useful  to address problems as renormalization and regularization in quantum field theory  and to address cosmological puzzles as the observed huge amounts of dark energy and dark matter, up to now not probed  at fundamental scales. Among these attempts, $f(R)$-gravity can be considered a paradigm by which it is possible to preserve several good and well-established results of General Relativity,  without imposing "a priori" the form of the gravitational action, chosen to be $f(R)=R$ in the  Hilbert-Einstein's case. For comprehensive reviews on the argument, see e.g. \cite{Schmidt:2004,OdiRev,GRGrew,faraonirev}. From an epistemological viewpoint, considering $f(R)$-gravity is a statement of "ignorance": we do not know the "true" action of gravity but we know that it has to be constructed by  curvature invariants (the simplest is the Ricci scalar $R$) which well represents the space-time dynamics.   Observations and experiments could, in principle, help to reconstruct the effective form of the theory by solving inconsistencies and shortcomings at  various scales.

However, any theory of physics is  viable if initial value problem is correctly formulated. As a consequence, the following dynamical evolution is uniquely determined and  agrees with causality requests. Besides, also if the initial
value problem is  well-formulated, we need  further properties   which are: $i)$ small  perturbations in the initial data have to produce small perturbations in the subsequent dynamics over all the space-time where it is defined; $ii)$ changes in the initial data region have to preserve the causal structure
of the theory. If both these requirements are satisfied, the initial value problem of the theory is  also well-posed.

General Relativity has a well-formulated and well-posed initial value problem  and then it results a  stable theory
with a robust causal structure \cite{yvonne,yvonne2,Synge,Wald}.  Such a theoretical structure should be achieved also for  $f(R)$-gravity, if one wants to consider it a viable  extension of the Einstein theory.

The debate on the self-consistency of $f(R)$-gravity cannot leave  out of consideration the Cauchy problem   which gives some  concerns  coming from  higher-order derivatives in the field equations and the increased number of degrees of freedom. For example, it is, up to now, unclear if the prolongation of standard methods can be used in
order to tackle this problem for  any $f(R)$-theory. Hence it is doubtful that the Cauchy problem could be properly addressed, if one simply takes into account the results already obtained for fourth-order theories
stemming from  quadratic Lagrangians \cite{Teyssandier:Tourrenc:1983,Kerner}.

On the other hand, being $f(R)$-gravity, like General Relativity, a gauge theory, the initial value formulation could depend on suitable constraints and on suitable "gauge choices" that mean a choice of coordinates so that the Cauchy problem results well-formulated and, possibly, well-posed. In \cite{Teyssandier:Tourrenc:1983,Noakes},
the initial value problem was studied for quadratic Lagrangians in the metric approach with the conclusion that it is well-posed. On the other hand, in \cite{Faraoni}, the Cauchy problem for generic $f(R)$-models has been studied in metric and Palatini approaches using the dynamical equivalence between these theories and the
Brans-Dicke gravity. The result was that the problem is well-formulated for metric approach in presence of matter and also well-posed in vacuo. For the Palatini approach, instead, the Cauchy problem is not well-formulated and well-posed since, considering the $3+1$ ADM formulation of the equivalent scalar-tensor theory, the  Brans-Dicke parameter $\omega=-3/2$ leads to a non-dynamical  field $\phi$ and to the impossibility of a first-order formulation of the problem.

A different approach is adopted in \cite{noi}: it is possible to show that the Cauchy problem of metric-affine
$f(R)$-gravity is well-formulated and well-posed in vacuo, while it can be, at least, well-formulated for various form of matter fields. The reason of the apparent contradiction with respect to the results in \cite{Faraoni} lies on the above mentioned gauge choice. Following \cite{Synge}, Gaussian normal coordinates have been adopted. Such a choice, introducing further constraints on the Cauchy data surface, results more suitable to set the initial
value problem in such a way that the well-formulation can be easily achieved. As a general remark, we can say that the well-position cannot be achieved  for any metric-affine $f(R)$-gravity theory but it has to be formulated specifying, case by case, the source term in the field equations.
However, it is straightforward to demonstrate that, in vacuum case, as well as for electromagnetic and generic Yang-Mills fields acting as sources, the Cauchy problem results always well-formulated and well-posed since it is possible to demonstrate that  $f(R)$-gravity reduces to $R+\Lambda$, that is the General Relativity plus  a cosmological constant \cite{noi}.

In this paper, we want to face the Cauchy problem for  metric-affine $f(R)$-gravity, in the Palatini approach and with torsion, assuming  perfect-fluid matter as source. Performing the conformal transformation from the Jordan to the Einstein frame and following the approach by Bruhat adopted for General Relativity \cite{yvonne,yvonne2}, it is possible to show that the initial value problem result well-formulated and well-posed in this case.
The layout of the paper is the following. In Sec. II, we derive the field equations for a generic $f(R)$-gravity theory discussing the differences and the analogies between the approach  {\it \`a la\/} Palatini and with torsion. Both  theories can be dealt under the same standard by defining a suitable scalar field. A main role, as we will show, is played by the conservation laws. Sec.III is devoted to the conformal transformation from the Jordan frame to the Einstein frame while the well--posedness of the Cauchy problem in presence of perfect-fluid matter is discussed in Sec.IV. The key of the demonstration is due to the fact that it is possible to achieve a perfect-fluid form for the stress-energy tensor in both frames and the conservation laws are preserved under the conformal transformation. This fact, as we will show, allows the well--posedness. The relevant example $f(R)=R+\alpha R^2$ is discussed in Sec.V. Conclusions are drawn in Sec. VI.

\section{The field equations of $f(R)$-gravity in the metric- affine formulation}
In the metric--affine formulation of $f(R)$-gravity, the (gravitational) dynamical fields are pairs $(g,\Gamma)\/$ consisting of a pseudo--Riemannian metric $g\/$ and a linear connection $\Gamma\/$ on the space-time manifold $M\/$ \cite{francaviglia}. In the Palatini approach, the connection $\Gamma\/$ is torsionless but it is not requested to be metric--compatible, instead, in the approach with torsion, the dynamical connection $\Gamma$ is forced to be metric but with torsion different from zero.

The field equations are derived from an action functional of the form
\begin{equation}\label{2.1}
{\cal A}\/(g,\Gamma)=\int{\left(\sqrt{|g|}f\/(R) + {\cal L}_m\right)\,ds}
\end{equation}
where $f(R)$ is a real function, $R\/(g,\Gamma) = g^{ij}R_{ij}\/$ (with $R_{ij}:= R^h_{\;\;ihj}\/$) is the scalar curvature associated with the connection $\Gamma\/$ and ${\cal L}_m\,ds\/$ is a suitable material Lagrangian.

Assuming that the material Lagrangian does not depend on the dynamical connection, the field equations are
\begin{subequations}\label{2.2}
\begin{equation}\label{2.2a}
f'\/(R)R_{ij} - \frac{1}{2}f\/(R)g_{ij}=\Sigma_{ij}\,,
\end{equation}
\begin{equation}\label{2.2b}
T_{ij}^{\;\;\;h} = - \frac{1}{2f'\/(R)}\de{f'\/(R)}/de{x^p}\/\left(\delta^p_i\delta^h_j - \delta^p_j\delta^h_i\right)
\end{equation}
\end{subequations}
for $f(R)$-gravity with torsion, and
\begin{subequations}\label{2.3}
\begin{equation}\label{2.3a}
f'\/(R)R_{ij} - \frac{1}{2}f\/(R)g_{ij}=\Sigma_{ij}\,,
\end{equation}
\begin{equation}\label{2.3b}
\nabla_k\/(f'(R)g_{ij})=0\,,
\end{equation}
\end{subequations}
for $f(R)$-gravity in the  Palatini approach \cite{CCSV1}. In Eqs.
\eqref{2.2a} and \eqref{2.3a}, the quantity ${\displaystyle
\Sigma_{ij}:= -
\frac{1}{\sqrt{|g|}}\frac{\delta{\cal L}_m}{\delta g^{ij}}\/}$ plays the role
of stress-energy tensor, the source of the field equations.
\\
Considering the trace of Eqs.
\eqref{2.2a} and \eqref{2.3a}, we obtain a relation between the
curvature scalar $R\/$ and the trace of the stress-energy tensor
$\Sigma:=g^{ij}\Sigma_{ij}\/$. Indeed, we have
\begin{equation}\label{2.4}
f'\/(R)R -2f\/(R) = \Sigma\,.
\end{equation}
From now on, we shall suppose that the relation \eqref{2.4} is invertible as well as that $\Sigma\not=\const\/$ (this implies, for example, $f(R)\/$ different from $\alpha R^2\/$ which is only compatible with $\Sigma=0\/$) . Under these hypotheses, the curvature scalar $R\/$ can be expressed as a suitable function of $\Sigma\/$, namely
\begin{equation}\label{2.5}
R=F(\Sigma)\,.
\end{equation}
If $\Sigma=\const$, General Relativity plus the cosmological constant is recovered \cite{CCSV1}.
Starting from eq. \eqref{2.5} and defining the scalar field
\begin{equation}\label{2.6}
\varphi:=f'(F(\Sigma))
\end{equation}
we can put the Einstein--like field equations of both {\it \`a la\/} Palatini and with torsion theories in the same form \cite{CCSV1,Olmo}, that is
\begin{equation}\label{2.7}
\begin{split}
\tilde{R}_{ij} -\frac{1}{2}\tilde{R}g_{ij}= \frac{1}{\varphi}\Sigma_{ij}
+ \frac{1}{\varphi^2}\left( - \frac{3}{2}\de\varphi/de{x^i}\de\varphi/de{x^j}
+ \varphi\tilde{\nabla}_{j}\de\varphi/de{x^i} + \frac{3}{4}\de\varphi/de{x^h}\de\varphi/de{x^k}g^{hk}g_{ij} \right. \\
\left. - \varphi\tilde{\nabla}^h\de\varphi/de{x^h}g_{ij} -
V\/(\varphi)g_{ij} \right)\,,
\end{split}
\end{equation}
where we have introduced the effective potential
\bigskip\noindent
\begin{equation}\label{2.8}
V\/(\varphi):= \frac{1}{4}\left[ \varphi
F^{-1}\/((f')^{-1}\/(\varphi)) +
\varphi^2\/(f')^{-1}\/(\varphi)\right]\,,
\end{equation}
for the scalar field $\varphi\/$. In Eq. \eqref{2.7}, $\tilde{R}_{ij}\/$, $\tilde{R}\/$ and $\tilde\nabla\/$ denote respectively the Ricci tensor, the scalar curvature and the covariant derivative associated with the Levi--Civita connection of the dynamical metric $g_{ij}$.
\\
Therefore, if the dynamical connection $\Gamma\/$ is not coupled with matter, both the theories (with torsion and Palatini--like) generate identical Einstein--like field equations. On the contrary, the field equations for the dynamical connection are different and (in general) give rise to different solutions. In fact, the connection $\Gamma$ solution of eqs. \eqref{2.2b} is
\begin{equation}\label{2.9}
\Gamma_{ij}^{\;\;\;h} =\tilde{\Gamma}_{ij}^{\;\;\;h} + \frac{1}{2\varphi}\de\varphi/de{x^j}\delta^h_i - \frac{1}{2\varphi}\de\varphi/de{x^p}g^{ph}g_{ij}
\end{equation}
where $\tilde{\Gamma}_{ij}^{\;\;\;h}\/$ denote the coefficients of the Levi--Civita connection associated with the metric $g_{ij}$, while the connection $\bar\Gamma\/$ solution of eqs. \eqref{2.3b} is
\begin{equation}\label{2.10}
\bar{\Gamma}_{ij}^{\;\;\;h}= \tilde{\Gamma}_{ij}^{\;\;\;h} +
\frac{1}{2\varphi}\de\varphi/de{x^j}\delta^h_i -
\frac{1}{2\varphi}\de\varphi/de{x^p}g^{ph}g_{ij} +
\frac{1}{2\varphi}\de\varphi/de{x^i}\delta^h_j\,,
\end{equation}
and coincides with the Levi--Civita connection induced by the conformal metric $\bar{g}_{ij}:=\varphi g_{ij}$. By comparison, the connections $\Gamma$ and $\bar\Gamma$ satisfy the relation
\begin{equation}\label{2.11}
\bar{\Gamma}_{ij}^{\;\;\;h} =\Gamma_{ij}^{\;\;\;h} +
\frac{1}{2\varphi}\de\varphi/de{x^i}\delta^h_j\,.
\end{equation}
Of course, the Einstein--like equations \eqref{2.7} are coupled with the matter field equations. In this respect, it is worth pointing out that eqs. \eqref{2.7} imply the same conservation laws holding in General Relativity. We have, in fact,
\begin{Proposition}\label{ProB.1}
Eqs. \eqref{2.6}, \eqref{2.7} and \eqref{2.8} imply the standard
conservation laws $\tilde{\nabla}^j\/\Sigma_{ij}=0$
\end{Proposition}
\demo First of all, we recall that Eq. \eqref{2.6} and \eqref{2.8}
are equivalent to the relation
\begin{equation}\label{B.7bis}
\Sigma -\frac{6}{\varphi}V(\varphi) + 2V'(\varphi)=0
\end{equation}
(see \cite{CCSV1} for the proof). After that, taking the trace of
Eq. \eqref{2.7} into account, we get
\begin{equation}\label{B.8}
\Sigma= -\varphi\tilde{R} -
\frac{3}{2}\frac{1}{\varphi}\varphi_i\varphi^i +
3\tilde{\nabla}_i\varphi^i + \frac{4}{\varphi}V(\varphi)
\end{equation}
where,  for the sake of  simplicity, we have defined ${\displaystyle \varphi_i :=
\de\varphi/de{x^i}}$. Substituting Eq. \eqref{B.8} in Eq.
\eqref{B.7bis}, we obtain
\begin{equation}\label{B.9}
\tilde{R} + \frac{3}{2}\frac{1}{\varphi^2}\varphi_i\varphi^i -
\frac{3}{\varphi}\tilde{\nabla}_i\varphi^i +
\frac{2}{\varphi^2}V(\varphi) - \frac{2}{\varphi}V'(\varphi)=0\,.
\end{equation}
We rewrite Eq. \eqref{2.7} in the form
\begin{equation}\label{B.10}
\begin{split}
\varphi\tilde{R}_{ij} -\frac{\varphi}{2}\tilde{R}g_{ij}= \Sigma_{ij} + \frac{1}{\varphi}\left( - \frac{3}{2}\varphi_i\varphi_j + \varphi\tilde{\nabla}_{j}\varphi_i + \frac{3}{4}\varphi_h\varphi^h\/g_{ij} + \right. \\
\left. - \varphi\tilde{\nabla}^h\varphi_h\/g_{ij} -
V\/(\varphi)g_{ij} \right)\,.
\end{split}
\end{equation}
The covariant divergence of \eqref{B.10} yields
\begin{equation}\label{B.11}
\begin{split}
(\tilde\nabla^j\varphi)\tilde{R}_{ij} + \varphi\tilde\nabla^j\tilde{G}_{ij} -\frac{1}{2}\tilde{R}\tilde\nabla_i\varphi = \tilde\nabla^j\Sigma_{ij} + \left(\tilde\nabla^j\tilde{\nabla}_{j}\tilde\nabla_i - \tilde\nabla_i\tilde{\nabla}^j\tilde\nabla_j\right)\varphi +\\
+
\tilde\nabla^j\left[\frac{1}{\varphi}\left(-\frac{3}{2}\varphi_i\varphi_j
+ \frac{3}{4}\varphi_h\varphi^h\/g_{ij} -
V\/(\varphi)g_{ij}\right)\right]\,.
\end{split}
\end{equation}
By definition, the Einstein and the Ricci tensors satisfy
$\tilde\nabla^j\tilde{G}_{ij}=0$ and
$(\tilde\nabla^j\varphi)\tilde{R}_{ij} =
\left(\tilde\nabla^j\tilde{\nabla}_{j}\tilde\nabla_i -
\tilde\nabla_i\tilde{\nabla}^j\tilde\nabla_j\right)\varphi$. Then
Eq. \eqref{B.11} reduces to
\begin{equation}\label{B.12}
-\frac{1}{2}\tilde{R}\tilde\nabla_i\varphi =
\tilde\nabla^j\Sigma_{ij}+
\tilde\nabla^j\left[\frac{1}{\varphi}\left(-\frac{3}{2}\varphi_i\varphi_j
+ \frac{3}{4}\varphi_h\varphi^h\/g_{ij} -
V\/(\varphi)g_{ij}\right)\right]
\end{equation}
Finally, making use of Eq. \eqref{B.9} it is easily seen that
\begin{equation}\label{B.13}
-\frac{1}{2}\tilde{R}\tilde\nabla_i\varphi =
\tilde\nabla^j\left[\frac{1}{\varphi}\left(-\frac{3}{2}\varphi_i\varphi_j
+ \frac{3}{4}\varphi_h\varphi^h\/g_{ij} -
V\/(\varphi)g_{ij}\right)\right]
\end{equation}
from which the conclusion $\tilde\nabla^j\Sigma_{ij}=0$ follows.
\enddemo
This result will be particularly useful for the following considerations.
\section{From the Jordan to the Einstein frame}
The field equations \eqref{2.7} can be simplified by passing from the Jordan to the Einstein frame.
Indeed, performing the conformal transformation $\bar{g}_{ij}=\varphi\/g_{ij}$, eqs. \eqref{2.7} assume the simpler
form (see for example \cite{CCSV1,Olmo})
\begin{equation}\label{3.1}
\bar{R}_{ij} - \frac{1}{2}\bar{R}\bar{g}_{ij} =
\frac{1}{\varphi}\Sigma_{ij} -
\frac{1}{\varphi^3}V\/(\varphi)\bar{g}_{ij}
\end{equation}
where $\bar{R}_{ij}\/$ and $\bar{R}\/$ are respectively the Ricci
tensor and the curvature scalar derived from the conformal metric
$\bar{g}_{ij}\/$. It is worth noticing that the conformal
transformation is working if the trace $\Sigma$ of the
stress-energy tensor is independent of the metric $g_{ij}$. Only in
this case in fact, eqs. \eqref{3.1} depend exclusively on the conformal
metric $\bar{g}_{ij}$ and the other matter fields.
\\
Again, eqs. \eqref{3.1} have to be considered together with the matter field equations. The latter are usually written in the Jordan frame, so involving the connection $\tilde\Gamma\/$ (or, also, $\Gamma\/$). Then, making use of the relations \eqref{2.10} and \eqref{2.11}, one can easily express the matter field equations in terms of the connection $\bar\Gamma\/$ instead of $\tilde\Gamma\/$ (or $\Gamma\/$).
\\
Moreover, for further applications, it is useful to show the  relations existing between the conservation laws  in the Jordan frame and in the Einstein frames. To this end, defining
\begin{equation}\label{3.2}
T_{ij}:=\frac{1}{\varphi}\Sigma_{ij} - \frac{1}{\varphi^3}V\/(\varphi)\bar{g}_{ij}
\end{equation}
the quantity appearing in the right--hand side of eqs. \eqref{3.1}, we can state the following
\begin{Proposition}\label{ProB.2}
Given the Levi-Civita connection $\bar\Gamma\/$,
derived from the conformal metric tensor $\bar{g}\/$,
and given the associated covariant derivative $\bar\nabla\/$,
the condition $\bar{\nabla}^j\/T_{ij}=0\/$ is equivalent to the
condition $\tilde{\nabla}^j\/\Sigma_{ij}=0$
\end{Proposition}
\demo
Let us develop the divergence
\begin{equation}\label{B.3}
\begin{split}
\bar{\nabla}^j\/T_{ij}= \frac{1}{\varphi}g^{sj}\bar{\nabla}_s\/T_{ij}= \frac{1}{\varphi}g^{sj}\left[\tilde{\nabla}_s\/T_{ij} - \frac{1}{2\varphi}\left(\de{\varphi}/de{x^i}\delta^q_s + \de{\varphi}/de{x^s}\delta^q_i - \de{\varphi}/de{x^u}g^{uq}g_{si}\right)T_{qj} +\right. \\
\left. - \frac{1}{2\varphi}\left(\de{\varphi}/de{x^j}\delta^q_s +
\de{\varphi}/de{x^s}\delta^q_j -
\de{\varphi}/de{x^u}g^{uq}g_{sj}\right)T_{iq}\right]\,.
\end{split}
\end{equation}
We have separately
\begin{equation}\label{B.4}
\begin{split}
\frac{1}{\varphi}g^{sj}\tilde{\nabla}_s\/T_{ij}=\frac{1}{\varphi}g^{sj}\tilde{\nabla}_s\/\left(\frac{1}{\varphi}\Sigma_{ij} - \frac{1}{\varphi^2}V\/(\varphi)g_{ij}\right)=\frac{1}{\varphi^2}\tilde{\nabla}^j\Sigma_{ij} - \frac{1}{\varphi^3}\de{\varphi}/de{x^s}\Sigma_i^{\;s} +\\
- \frac{1}{\varphi}\de
/de{x^s}\left(\frac{1}{\varphi^2}V\/(\varphi)\right)\delta^s_i
\end{split}
\end{equation}
\begin{equation}\label{B.5}
\begin{split}
\frac{1}{\varphi}g^{sj}\frac{1}{2\varphi}\left(\de{\varphi}/de{x^i}\delta^q_s + \de{\varphi}/de{x^s}\delta^q_i - \de{\varphi}/de{x^u}g^{uq}g_{si}\right)T_{qj}=\\
=\frac{1}{\varphi}g^{sj}\frac{1}{2\varphi}\left(\de{\varphi}/de{x^i}\delta^q_s + \de{\varphi}/de{x^s}\delta^q_i - \de{\varphi}/de{x^u}g^{uq}g_{si}\right)\/\left(\frac{1}{\varphi}\Sigma_{qj} - \frac{1}{\varphi^2}V\/(\varphi)g_{qj}\right)=\\
=\frac{1}{2\varphi^3}g^{sj}\/\left(\de{\varphi}/de{x^i}\Sigma_{sj} + \de{\varphi}/de{x^s}\Sigma_{ij} - \de{\varphi}/de{x^u}g_{si}\Sigma^u_{\;j}\right)+\\
-\frac{1}{2\varphi^4}g^{sj}\/\left(\de{\varphi}/de{x^i}V\/(\varphi)g_{sj} + \de{\varphi}/de{x^s}V\/(\varphi)g_{ij} - \de{\varphi}/de{x^u}V\/(\varphi)\delta^u_jg_{si}\right)=\\
=\frac{1}{2\varphi^3}\de{\varphi}/de{x^i}\Sigma -
\frac{2}{\varphi^4}\de{\varphi}/de{x^i}V\/(\varphi)
\end{split}
\end{equation}
\begin{equation}\label{B.6}
\begin{split}
\frac{1}{\varphi}g^{sj}\frac{1}{2\varphi}\left(\de{\varphi}/de{x^j}\delta^q_s + \de{\varphi}/de{x^s}\delta^q_j - \de{\varphi}/de{x^u}g^{uq}g_{sj}\right)T_{iq}=\\
=\frac{1}{\varphi}g^{sj}\frac{1}{2\varphi}\left(\de{\varphi}/de{x^j}\delta^q_s + \de{\varphi}/de{x^s}\delta^q_j - \de{\varphi}/de{x^u}g^{uq}g_{sj}\right)\/\left(\frac{1}{\varphi}\Sigma_{iq} - \frac{1}{\varphi^2}V\/(\varphi)g_{iq}\right)=\\
=\frac{1}{2\varphi^3}g^{sj}\/\left(\de{\varphi}/de{x^j}\Sigma_{is} + \de{\varphi}/de{x^s}\Sigma_{ij} - \de{\varphi}/de{x^u}g_{sj}\Sigma^u_{\;i}\right)+\\
-\frac{1}{2\varphi^4}g^{sj}\/\left(\de{\varphi}/de{x^j}V\/(\varphi)g_{si} + \de{\varphi}/de{x^s}V\/(\varphi)g_{ij} - \de{\varphi}/de{x^u}V\/(\varphi)\delta^u_ig_{sj}\right)=\\
-\frac{1}{\varphi^3}\de{\varphi}/de{x^s}\Sigma^{\;s}_i +
\frac{1}{\varphi^4}\de{\varphi}/de{x^i}V\/(\varphi)
\end{split}
\end{equation}
Collecting eqs. \eqref{B.4}, \eqref{B.5} and \eqref{B.6}, we have
then
\begin{equation}\label{B.7}
\bar{\nabla}^j\/T_{ij}=\frac{1}{\varphi^2}\tilde{\nabla}^j\Sigma_{ij}
+ \frac{1}{\varphi^3}\de{\varphi}/de{x^i}\left[-\frac{1}{2}\Sigma +
\frac{3}{\varphi}V(\varphi) - V'(\varphi)\right]=
\frac{1}{\varphi^2}\tilde{\nabla}^j\Sigma_{ij}
\end{equation}
because the identity $-\frac{1}{2}\Sigma +
\frac{3}{\varphi}V(\varphi) - V'(\varphi)=0$ holds identically,
being equivalent to the definition $\varphi=f'(F(\Sigma))$
\cite{CCSV1}.
\enddemo
\noindent
From Proposition \ref{ProB.2}, we conclude that the quantity \eqref{3.2} plays the role of the effective stress--energy tensor for the conformally transformed theory (for a discussion on the covariant conservation of energy-momentum tensor in modified gravity, see also \cite{Koivisto}).

\section{The well--posedness of the Cauchy problem}
 Let us consider now the  $f(R)\/$-gravity, in the metric-affine formalism, coupled with  perfect-fluid matter acting as  source in the field equations. We are going to demonstrate that, in the Einstein frame, the analysis of the related Cauchy problem can be carried out by following the same arguments developed in \cite{yvonne,yvonne2}.
To see this point, we start with  looking  for a metric $g_{ij}\/$ of signature $(-+++)\/$ in the Jordan frame. The stress-energy tensor of perfect--fluid matter is of the form
\begin{subequations}\label{4.1}
\begin{equation}\label{4.1a}
\Sigma_{ij}=(\rho + p)\,U_iU_j + p\,g_{ij}
\end{equation}
giving rise to corresponding matter field equations
\begin{equation}\label{4.1b}
\tilde\nabla_j\Sigma^{ij}=0\,.
\end{equation}
\end{subequations}
In eqs. \eqref{4.1}, the scalars $\rho$ and $p$ denote, respectively, the matter--energy density and the pressure of the fluid, while $U_i$ indicates the four velocity of the fluid, satisfying the obvious condition $g^{ij}U_iU_j =-1\/$.
\\
After performing the conformal transformation $\bar{g}_{ij}=\varphi g_{ij}\/$, we can express the field equations in the Einstein frame as
\begin{subequations}\label{4.2}
\begin{equation}\label{4.2a}
\bar{R}_{ij} - \frac{1}{2}\bar{R}\bar{g}_{ij} = T_{ij}
\end{equation}
and
\begin{equation}\label{4.2b}
\bar{\nabla}_j T^{ij}=0\,,
\end{equation}
\end{subequations}
where
\begin{equation}\label{4.3}
T_{ij}=\frac{1}{\varphi}(\rho + p)\,U_iU_j + \left( \frac{p}{\varphi^2} - \frac{V(\varphi)}{\varphi^3} \right)\,\bar{g}_{ij}
\end{equation}
is the effective stress--energy tensor.
In view of Proposition \ref{ProB.2}, eqs. \eqref{4.2b} are equivalent to eqs. \eqref{4.1b}. As we shall see, this is a key point of our discussion, allowing us to apply to the present case the results achieved in \cite{yvonne,yvonne2}.
Moreover,  we shall suppose for the moment that the scalar field $\varphi$ is positive, that is $\varphi > 0\/$. The opposite case $\varphi <0\/$, differing from the former only for some technical aspects, will be briefly discussed below.
\\
Under the stated assumptions, the four--velocity of the fluid in the Einstein frame can be expressed as $\bar{U}_i = \sqrt{\varphi}U_i\/$. In view of this, the stress--energy tensor \eqref{4.3} can be rewritten in terms of the four velocity $\bar{U}_i\/$ as
\begin{equation}\label{4.4}
T_{ij}=\frac{1}{\varphi^2}(\rho + p)\,\bar{U}_i\bar{U}_j + \left( \frac{p}{\varphi^2} - \frac{V(\varphi)}{\varphi^3} \right)\,\bar{g}_{ij}
\end{equation}
Furthermore, introducing the effective mass--energy density
\begin{subequations}\label{4.5}
\begin{equation}\label{4.5a}
\bar{\rho}:= \frac{\rho}{\varphi^2} + \frac{V(\varphi)}{\varphi^3}\,,
\end{equation}
and the effective pressure
\begin{equation}\label{4.5b}
\bar{p}:= \frac{p}{\varphi^2} - \frac{V(\varphi)}{\varphi^3}\,,
\end{equation}
\end{subequations}
the stress--energy tensor \eqref{4.4} assumes the final standard form
\begin{equation}\label{4.6}
T_{ij}=\left( \bar\rho + \bar{p} \right)\,\bar{U}_i\bar{U}_j + \bar{p}\,\bar{g}_{ij}\,.
\end{equation}
It is worth noticing that, starting from an equation of state of the form $\rho=\rho(p)\/$ and assuming that the relation \eqref{4.5b} is invertible $(p=p(\bar{p}))$, by composition with eq. \eqref{4.5a} we derive an effective equation of state $\bar{\rho}=\bar{\rho}(\bar{p})\/$. In addition, we recall that the explicit expression of the scalar field $\varphi\/$, as well as of the potential $V(\varphi)\/$, are directly related with the particular form of the function $f(R)\/$. Then, the requirement of invertibility of the relation \eqref{4.5b} together with the condition $\varphi >0\/$ (or, equivalently, $\varphi <0\/$) become criteria for the viability of the functions $f(R)\/$. In other words, they provide us with precise rules of selection for the admissible functions $f(R)\/$.
\\
From now on, the treatment of the Cauchy problem can proceed step by step as in \cite{yvonne,yvonne2}. We do not repeat Bruhat's analysis here, referring the reader to her well known papers. We only recall the conclusion stated in \cite{yvonne,yvonne2}, where it is proved that the Cauchy problem for the system of differential equations \eqref{4.2}, with stress--energy tensor given by eq. \eqref{4.6} and equation of state $\bar{\rho}=\bar{\rho}(\bar{p})\/$, is well--posed if the condition
\begin{equation}\label{4.7}
\frac{d\bar{\rho}}{d\bar{p}}\geq 1\,,
\end{equation}
is satisfied. We stress that, in order to check the requirement \eqref{4.7}, we do not need to invert explicitly the relation \eqref{4.5b}, but more simply, we can verify
\begin{equation}\label{4.8}
\frac{d\bar{\rho}}{d\bar{p}}=\frac{d\bar{\rho}/dp}{d\bar{p}/dp}\geq 1
\end{equation}
directly from the expressions \eqref{4.5} and the equation of state $\rho=\rho(p)\/$. Once again, the condition \eqref{4.8}, depending on the peculiar expressions of $\varphi$ and $V(\varphi)\/$, is strictly related to the particular form of the function $f(R)\/$. Then, condition \eqref{4.8} represents a further criterion for the viability of  $f(R)\/$-models whose initial value problem is well formulated and well posed.
\\
For completeness, we outline the case $\varphi <0\/$. We still suppose that the signature of the metric in the Jordan frame is $(-+++)\/$. Therefore, the signature of the conformal metric will be $(+---)\/$ and the four velocity of the fluid in the Einstein frame will be $\bar{U}_i = \sqrt{-\varphi}U_i\/$.
\\
The effective stress--energy tensor is given now by
\begin{equation}\label{4.9}
T_{ij}=-\frac{1}{\varphi^2}(\rho + p)\,\bar{U}_i\bar{U}_j + \left( \frac{p}{\varphi^2} - \frac{V(\varphi)}{\varphi^3} \right)\,\bar{g}_{ij}= \left( \bar\rho + \bar{p} \right)\,\bar{U}_i\bar{U}_j - \bar{p}\,\bar{g}_{ij}\,,
\end{equation}
where we have introduced the quantities
\begin{subequations}\label{4.10}
\begin{equation}\label{4.10a}
\bar{\rho}:= -\frac{\rho}{\varphi^2} - \frac{V(\varphi)}{\varphi^3}\,,
\end{equation}
and
\begin{equation}\label{4.10b}
\bar{p}:= -\frac{p}{\varphi^2} + \frac{V(\varphi)}{\varphi^3}\,,
\end{equation}
\end{subequations}
representing, as above, the effective energy-density and the effective pressure. At this point, everything will proceed again as in \cite{yvonne,yvonne2}, except for a technical aspect. The  quantity ${\displaystyle r:=\bar{\rho}+\bar{p}=-\frac{\rho + p}{\varphi^2}\/}$  is now negative (if, as usual, $\rho\/$ and $p\/$ are assumed to be positive). Therefore, instead of using the function $\log(f^{-2}r)$ as in \cite{yvonne,yvonne2} (where $f$ denotes the index of the fluid \cite{Lichnerowicz}), we need to take into account $\log(-f^{-2}r)$. The reader can easily verify that, with this choice,  Bruhat's arguments apply equally well.

\section{An example}
As an illustrative example of the above demonstration
let us take into account the $f(R)= R+ \alpha{R^2}$ theory coupled with dust which is a particular case of perfect-fluid matter. The matter stress-energy tensor
in the Jordan frame is $\Sigma_{ij} = \rho U_i U_j$ being $p=0$. %
\\
From the trace of the field equations \eqref{2.2a}, we derive the relation
\begin{equation}\label{5.1}
(1+ 2\alpha R)R -2R - 2\alpha R^2 = -\rho \qquad
\longleftrightarrow \qquad R=\rho\,,
\end{equation}
so that the scalar field \eqref{2.6} becomes
\begin{equation}\label{5.2}
\varphi(\rho) = f'(R(\rho))= 1+2\alpha\rho
\end{equation}
For small values of the density $\rho<<1\/$ (for example, the present cosmological density) and values of $|\alpha|\/$ not comparable with $1/\rho\/$, we can reasonably suppose $\varphi>0\/$, independently of the sign of the parameter $\alpha\/$.
\\
Let us consider now the potential \eqref{2.8}
\begin{equation}\label{5.3}
V(\varphi)= \frac{1}{4}\left[ \varphi F^{-1}((f')^{-1}(\varphi)) +
\varphi^2(f')^{-1}(\varphi)\right]
\end{equation}
Being $(f')^{-1}(\varphi)=\rho$, one has
\begin{equation}\label{5.4}
\frac{1}{4}\varphi^2(f')^{-1}(\varphi)=\frac{1}{4}(1+2\alpha\rho)^2
\frac{1}{2\alpha}(1+2\alpha\rho
-1)=\frac{1}{4}(1+2\alpha\rho)^2\rho\,,
\end{equation}
and considering the relation $F^{-1}(Y)=f'(Y)K -2f(Y)$, it is
\begin{equation}\label{5.5}
\frac{1}{4}F^{-1}((f')^{-1}(\varphi))=\frac{1}{4}F^{-1}(\rho)=-\rho\,.
\end{equation}
We have also
\begin{equation}\label{5.6}
\frac{1}{4}\varphi F^{-1}((f')^{-1}(\varphi))=
-\frac{(1+2\alpha\rho)\rho}{4}\,,
\end{equation}
and then we conclude that
\begin{equation}\label{5.7}
V(\varphi(\rho))= \frac{\alpha\rho^2(1+2\alpha\rho)}{2}\,.
\end{equation}
In the Einstein frame, the stress--energy tensor is expressed as
\begin{equation}\label{5.8}
T_{ij}= \frac{\rho}{\varphi^2}\bar{U}_i\bar{U}_j -\frac{V(\varphi)}{\varphi^3}\bar{g}_{ij}\,.
\end{equation}
The latter can be seen as the stress--energy tensor of a perfect fluid with density and pressure given respectively by
\begin{subequations}\label{5.9}
\begin{equation}\label{5.9a}
\bar{\rho}:= \frac{\rho}{\varphi^2} + \frac{V(\varphi)}{\varphi^3}=\frac{2\rho + \alpha\rho^2}{2(1+2\alpha\rho)^2}\,,
\end{equation}
and
\begin{equation}\label{5.9b}
\bar{p}:= - \frac{V(\varphi)}{\varphi^3}=-\frac{\alpha\rho^2}{2(1+2\alpha\rho)^2}\,.
\end{equation}
\end{subequations}
Under the stated assumptions, the function \eqref{5.9b} is invertible, indeed, for $\rho>0\/$
\begin{equation}\label{5.10}
\frac{d\bar{p}}{d\rho}= -\frac{4\alpha\rho}{4\/(1+2\alpha\rho)^3}\not= 0\,.
\end{equation}
Moreover, we have
\begin{equation}\label{5.11}
\frac{d\bar{\rho}}{d\rho}=\frac{4-4\alpha\rho}{4\/(1+2\alpha\rho)^3}
\end{equation}
and then
\begin{equation}\label{5.12}
\frac{d\bar{\rho}}{d\bar{p}}=\frac{d\bar{\rho}/dp}{d\bar{p}/dp}=\frac{-1+\alpha\rho}{\alpha\rho}\geq 1 \quad\Longleftrightarrow\quad \alpha <0\,.
\end{equation}
This last result is selecting the form of the physically viable theories. It is relevant to note that  $\alpha$  negative defined allows stable cosmological solutions and positively defined massive states (see \cite{barrow} and references therein).

\section{Conclusions}
Following the prescriptions in \cite{yvonne,yvonne2}, it is possible to show that the Cauchy problem for metric-affine $f(R)$-theories {\it \`a la\/} Palatini and with torsion, in presence of perfect--fluid matter acting as  source, is well formulated and well posed. The key points of the demonstration are: $i)$ the conservation laws (the contracted Bianchi identities) are preserved under the conformal transformation from the Jordan to the Einstein frame; $ii)$ the Bruhat arguments can be applied if it is possible to recast the  stress--energy tensor in the perfect--fluid form, in both frames; $iii)$ the condition ${\displaystyle \frac{d\bar{\rho}}{d\bar{p}}\geq 1}$, specifically one of the Bruhat conditions, allows to select suitable $f(R)$-models and to formulate the energy theorems for this kind of theories \cite{magnano}.

It is worth noticing the role of the perfect--fluid which allows both the well formulation \cite{noi} as well as the well--posedness of the problem. Besides, it acts also as a sort of "selection rule" since not any $f(R)$-model is consistent with the well--posedness of the initial value problem but only those where the condition ${\displaystyle \frac{d\bar{\rho}}{d\bar{p}}\geq 1}$ holds, as we have seen for the $f(R)=R+\alpha R^2$ case.

As a concluding remark, we can say that the Cauchy problem result, in general, well formulated for $f(R)$-gravity in metric-affine formalism \cite{noi} as well as in metric formalism \cite{Faraoni}. On the other hand, the well--posedness, which always holds in vacuum (and also in the case of coupling with an electromagnetic field or with Yang--Mills fields) \cite{noi}, strictly depends on the source. In the case of perfect--fluid matter, it works, essentially,  because the problem can be reduced to the Einstein frame by a conformal transformation. The Cauchy problem for other forms of source will be the arguments of future investigations.

\end{document}